\documentclass{optica-article}

\journal{opticajournal} 

\articletype{Research Article}

\usepackage{lineno}

\usepackage{gensymb}
\usepackage{subcaption} 
\usepackage{siunitx}

\newcommand{\gt}{g^{(2)}}
\newcommand{\Gt}{G^{(2)}}

\usepackage[capitalize]{cleveref}
\crefname{section}{Sec.}{Secs.}
\Crefname{section}{Section}{Sections}
\Crefname{table}{Table}{Tables}
\crefname{table}{Tab.}{Tabs.}

\begin{document}

\title{Massively Multiplexed Wide-field Photon Correlation Sensing}

\author{Shay Elmalem,\authormark{1} Gur Lubin,\authormark{1,2,3} Michael Wayne, \authormark{4} Claudio Bruschini,\authormark{4} Edoardo Charbon\authormark{4} and Dan Oron\authormark{1,*}}

\address{\authormark{1}Department of Physics of Complex Systems, Weizmann Institute of Science, Rehovot, Israel\\
\authormark{2}JILA, National Institute of Standards and Technology and University of Colorado, Boulder, Colorado, USA\\
\authormark{3}Department of Physics, University of Colorado, Boulder, Colorado, USA\\
\authormark{4}School of Engineering, École polytechnique fédérale de Lausanne (EPFL), Neuchâtel, Switzerland}

\email{\authormark{*}dan.oron@weizmann.ac.il}

\begin{abstract*} 

Temporal photon correlations have been a crucial resource for quantum and quantum-enabled optical science for over half a century. However, attaining non-classical information through these correlations has typically been limited to a single point (or at best, a few points) at-a-time. We perform here a massively multiplexed wide-field photon correlation measurement using a large $500\times500$ single-photon avalanche diode array, the SwissSPAD3. We demonstrate the performance of this apparatus by acquiring wide-field photon correlation measurements of single-photon emitters, and illustrate two applications of the attained quantum information: wide-field emitter counting and quantum-enabled super-resolution imaging (by a factor of $\sqrt{2})$. The considerations and limitations of applying this technique in a practical context are discussed. Ultimately, the realization of massively multiplexed wide-field photon correlation measurements can accelerate quantum sensing protocols and quantum-enabled imaging techniques by orders of magnitude.

\end{abstract*}

\section{Introduction} 
\label{sec:01_intro}
The pivotal Hanbury Brown and Twiss (HBT) experiment \cite{HBT}, and insights it inspired on the quantum nature of light, led to various spin-offs and applications utilizing photon correlations and the photon bunching and anti-bunching effects. Such applications have been presented in the fields of astronomy \cite{dravins_stellarIntensityRev}, communications/computing \cite{gisin2007quantComm,couteau2023quanComAndComputing}, remote sensing \cite{lloyd2008enhanced, england2019quantum} and bio-imaging \cite{schwartz2012improved, schwartz2013superresolution, tenne2019super} to name a few. However, to date, most of these were based on either single/few-pixel single photon detectors or slow (and noisy) detector arrays such as EMCCDs. Over the past two decades, the performance of single-photon avalanche diode (SPAD) arrays has advanced considerably \cite{Charbon2013spadBookChapter, bruschini2019spadReview}. In recent years, large, high-performance SPAD array detectors have become available, offering a route for multiplexed photon correlation experiments. Specifically, the SwissSPAD3 (SS3) \cite{SS3} is a state-of-the-art $0.25 Mpix$ monolithic SPAD array, enabling wide-field measurements with single photon sensitivity. This sensor, integrated in a dedicated setup, is utilized to achieve massively multiplexed sensing of photon statistics. As presented in this work, measurement in such a setup provides both the conventional intensity information via photon counting and per-pixel photon statistics; the fusion of the intensity and photon correlation information holds potential for advanced imaging applications, such as quantum super-resolution (SR) imaging and quantum-assisted localization microscopy.

Photon statistics have proven to be a powerful tool across various fields. While the original HBT experiment utilized classical intensity correlations \cite{HBT}, its quantum-interpretation \cite{Glauber_1963_a, Glauber_1963_b} laid the foundations for a new field of research \cite{Kimble_1977_antBunch,Aspect_1981_Bell}. In astronomy, it enabled high-resolution measurements of stellar diameters and the study of astronomical objects \cite{dravins2012stellar, dravins_stellarIntensityRev}. Expanding to quantum communications, photon correlations have been used to enhance security protocols and enable quantum key distribution \cite{gisin2007quantComm, scarani2009security}. In remote sensing applications, quantum lidar systems have utilized photon correlation techniques to achieve enhanced sensitivity and resolution compared to classical systems \cite{lloyd2008enhanced, england2019quantum}. In bio-imaging and spectroscopy, photon correlation had been presented as an additional resource of information that can provide or assist in SR imaging \cite{schwartz2013superresolution, tenne2019super, lubin2022photon}. These examples among many others, demonstrate the versatility of photon correlation methods derived from the principles established by the HBT experiment.

Traditional implementations have mostly been limited by the reliance on single/few-pixel single-photon detectors. The ongoing development of SPAD array technology, growing from modest-sized arrays (tens to hundreds of pixels) to mega-pixel sized arrays, is enabling larger scale applications \cite{Charbon2013spadBookChapter,bruschini2019spadReview, dutton_SPAD_arr,perenzoni_SPAD_arr,gyongy_SPAD_arr,ATLAS_1,ATLAS_2}. One of the leading vectors of this trend is the SwissSPAD family; advances in the CMOS technology enabled the introduction of SPAD arrays in a camera-like form, starting with the $512\times128$ pixels SwissSPAD1 \cite{SS1} and the following $512\times512$ pixels SwissSPAD2 \cite{SS2}. Recent research has explored the use of these sensors in various imaging modalities; the advantages of such SPAD arrays have been presented for fluorescence lifetime imaging (FLIM) \cite{Xavier_FLIM,Faccio_FLIM}, quantum entanglement based imaging \cite{defienne2021full}, 3D and plenoptic imaging \cite{Gupta_3D,Dangelo_CPI,Dangelo_CPI_rev}, LIDAR \cite{Faccio_SPAD_lidar}, computational imaging \cite{Faccio_compImg, Gupta_burst}, and even software-defined camera emulation \cite{Gupta_SodaCam}, to name a few. Despite these developments, the integration of per-pixel photon statistics for imaging and sensing applications is still in its early stages. 

Recently, the SS3 \cite{SS3} has been introduced, featuring a $500\times500$ pixels array with $\SI{16.38}{\micro m}$ pixel pitch, fill-factor of $10.5\%$ (without micro-lenses) and maximum photon detection probability (PDP) of $\mathord{\sim}50\%$ at $\SI{520}{nm}$ (and over $30\%$ across the entire visible spectrum). The SS3 enables wide-field measurements with per-pixel single-photon sensitivity and practically noiseless readout, at maximal frame rate of $\SI{97.7}{kframes/s}$. For advanced application, it can support high-temporal-resolution dual time-gating. As explored in this work, this performance facilitates massively multiplexed sensing of photon statistics, providing both conventional intensity images and detailed photon correlation data at each pixel. Such capabilities can be utilized for advanced imaging and sensing applications, even in photon-starved scenarios, as presented and discussed in this work.

\section{Method}
\label{sec:02_method}

The proposed setup is based on a standard wide-field epifluorescence microscope, with a sub-pico-second laser as an illumination source, and an SS3 detector as the sensing device. Photon statistics sensing is demonstrated by measuring a sample of dispersed quantum dots (QDs), a prevalent sub-Poissonian photon source. By adequately setting the illumination laser properties with respect to the SS3 parameters (as detailed in the following sub-section), the QDs act as single-photon emitters in each acquired frame of the SS3. The acquired image stack $I_0(x,y,t)$ (where $1<x,y<500$ and $1<t<T$), contains $T$ binary frames of size $500\times500$. Each frame represents single-photon detection (or no detection) in each pixel at time $t$. $I_0$ is then used to calculate photon statistics, and can also be aggregated (fully or partially) to generate an intensity image or a video stream.

\subsection{Setup overview}
\begin{figure}[htbp]
\centering\includegraphics[width=7cm]{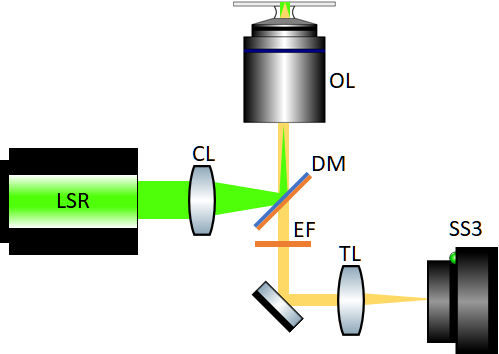}
\caption{{\bf{Setup overview.}} The illumination laser (LSR) is focused using the condenser lens (CL) and directed by the dichroic mirror (DM) to the back focal plane of the objective lens (OL). The light is collimated by the OL to achieve wide-field illumination. The fluorescent light is then collected by the OL, filtered by the the DM and emission filter (EF), and focused using the tube lens (TL) on the sensor plane of the SwissSPAD3 (SS3).}
\label{fig:setup}
\end{figure}

The setup (see \cref{fig:setup}) is based on a standard inverted microscope (Zeiss Axiovert 35). A $\SI{300}{fs}$ pulsed laser (Spirit One, SpOne-8-F2P-SHG by Spectra Physics) at a central wavelength of $\SI{520}{nm}$ is used as an illumination source. The light is reflected using a dichroic mirror (Semrock FF552-Di02), and focused on the back focal plane of an infinity-corrected, oil-immersed objective lens (Zeiss $63\times$ $NA=1.4$); the objective is collimating the light to the image plane, resulting in wide-field illumination. The sample is made of an ensamble of QDs, utilized as fluorescent single-photon emitters. The emission is then collected using the same objective, filtered using the dichroic mirror and an emission filter (Semrock BLP01-532R-25), and imaged on the SS3 by the internal tube lens of the microscope. 

The basic principle allowing this scheme is the utilization of the QDs as single-photon emitters (as previously presented in similar settings \cite{schwartz2013superresolution,tenne2019super}). This is achieved by a proper illumination setting: the laser pulse width ($t_p=\SI{300}{fs}$) is considerably shorter compared to the QDs radiative lifetime ($\tau_{QD}\geq\SI{50}{ns}$). This practically eliminates the possibility of re-excitation following emission within a single excitation pulse. Therefore, in such a setting the QDs behave as relatively pure single-photon emitters per illumination pulse. The emitted photons are then collected using the objective and imaged on the SS3. The total measured magnification is $M=110$, therefore the diameter of a diffraction-limited spot (i.e.\ the full Airy disk size) in the image plane is $d=M\frac{1.22\lambda}{NA}=\SI{57}{\micro m}$.

The image is captured using the SS3 (with its pixel pitch of $\SI{16.38}{\micro m}$), therefore, each diffraction-limited spot is mapped into $\mathord{\sim}3x3$ pixels. The SS3 native frame-rate is $f_{SS3}=\SI{97.7}{kframes/s}$. Ideally, to achieve one illumination pulse per frame, the laser repetition rate should be synced to the frame acquisition rate, or set asynchronously to $f_{LSR}=f_{SS3}-\epsilon$ (as performed in \cite{schwartz2013superresolution}). Due to firmware issues in synchronizing the laser and the SS3, the laser is set to a marginally higher repetition rate of $f_{LSR}=\SI{100}{kHz}$. This results in a negligible bias, as discussed in the appendix.

\subsection{Second-order correlation calculation}
\begin{figure}[htbp]
    \centering
    \begin{subfigure}{0.45\textwidth}
        \centering
        \includegraphics[width=\linewidth]{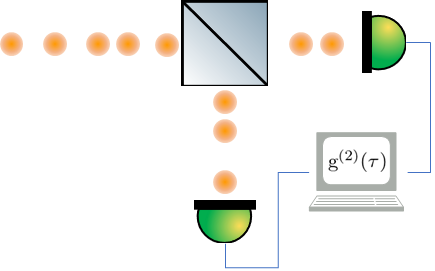} 
        \caption{A standard HBT setup} 
        \label{fig:HBT_orig}
    \end{subfigure}
    \hfill 
    \begin{subfigure}{0.3\textwidth}
        \centering
        \includegraphics[width=\linewidth]{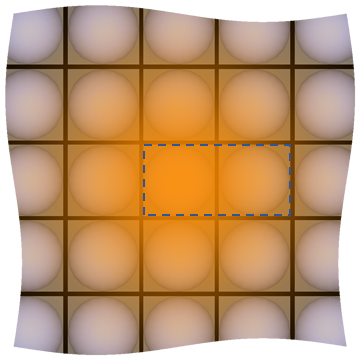} 
        \caption{Diffraction spot as a natural beam-splitter on the pixel grid} 
        \label{fig:pixOnGrid}
    \end{subfigure}
    
    \caption{Standard two-detector HBT vs. proposed HBT setting: in the standard (left) HBT setup, a beam splitter is used to split the light between two photo-detectors, whose output is used to calculate the photon correlations. Utilizing the same principle, in the proposed (right) setup, the diffraction spot replaces the beam splitter, by spreading the input light over several pixels of the SPAD array. Each pixel pair (for example, the central and its right neighbor, enclosed in the dashed line) are considered as pair of detectors in an HBT experiment.}
    \label{fig:HBT_and_pixGrid}
\end{figure}

The acquired binary frames are used to calculate the second-order correlation function $\gt(x,y,\tau)$. Light diffraction serves as a natural beam-splitter (see \cref{fig:HBT_and_pixGrid}), where light emitted from a point source in the sample is illuminating several adjacent pixels in the detector. Thus, each pair of adjacent pixels can be treated as an independent HBT experiment. By considering edge-sharing neighbors, (almost) each array pixel has four neighbors, thus the setup realizes $\mathord{\sim} 4 \cdot 500 \cdot 500 = 1 \cdot 10^6$ parallel HBT experiments. In the current setting, where the diffraction-limited spot is mapped to a $\mathord{\sim}3x3$ pixels area, most of the photon-pairs signal is contained in the four edge-sharing neighbors. However, with a larger magnification (or denser over-sampling) setting, also the corner-sharing neighbors, or even second-degree neighbors can be considered to enhance the SNR (the number of neighbors to include is a design consideration which mainly depends on the over-all photon detection probability $P_d$ and the SNR, as discussed in \cref{sec:04_discussion} and the appendix).

The un-normalized second-order correlation function of each adjacent pixel pair, $\Gt_n(x,y,\tau)$, is calculated by: 
\begin{equation}
    \Gt_n(x,y,\tau) =  \sum_{t=0}^{T} I_0(x,y,t)\cdot I_n(x,y,t-\tau),
    \label{eq:G(2)_n}
\end{equation}
where $T$ is the number of binary images, $I_0(x,y,t)$ is the image stack and $I_n$ is the spatially shifted image stack, according to the selected neighbor $n$ (for example, starting from the right neighbor and rotating counter clockwise- $I_1(x,y)=I(x-1,y); I_2(x,y)=I(x-1,y-1)$ and so on). The \emph{normalized} second-order correlation function, $\gt(x,y,\tau)$, is calculated by:
\begin{equation}
    \gt(x,y,\tau) =  \frac{\Gt(x,y,\tau)}{\Gt(x,y,\infty)},
    \label{eq:g(2)}
\end{equation}
where $\Gt(x,y,\infty)$ is the second-order correlation at `infinite' time delay (approximated by averaging $\Gt(x,y,\tau)$ over a domain of large values of $\tau$, as defined in \cref{sec:02_sub_pipeline}, \cref{eq:Gt_inf}).

\subsection{Sensitivity and error factors} 
When sensing second-order correlations, the required signal is photon pairs. To avoid saturation-related issues, typical photon-correlation experiments employ a single-photon detection probability of $P_d<<1$ (per-pulse, or per-frame in the current scheme), leading to a $(P_d)^2<<P_d$ probability to detect a photon pair. As this is a relatively infrequent event, special attention to the error factors is required, and their compensation is a crucial part of the data processing pipeline. In the proposed scheme, two inherent characteristics of large pixelated sensors are compensated for: crosstalk (CT) and dark-counts (DC). The corrections and calibrations presented below are an extension of the established methods demonstrated for smaller SPAD arrays\cite{Gur_quantCorr_Spad23}.

CT is the phenomenon in which a detection in one pixel generates a false detection in one of its neighbors (either optically or electronically). While this is mostly a negligible issue in intensity imaging, it is a crucial problem when considering second-order correlation, as CT generates false photon pairs for the same frame (i.e.\ for $\gt(0)$). In \cite{SS3}, the probability of a CT event $P_{CT}$ in the SS3 had been estimated globally, using a dark measurement and by inspecting the values near hot pixels. While this is a suitable method to get an overall estimation for $P_{CT}$, using this global value to compensate the desired second order correlation measurements did not lead to satisfying results; careful analysis shows that $P_{CT}$ appears to be inhomogeneous, and also somewhat anisotropic with respect to the detector axes (note that anisotropy also exists in the values reported in \cite{SS3}). Therefore, to estimate a $P^{CT}_n(x,y)$ map for each neighboring pixel pair (denoted by $n$), a $\SI{200}{s}$ measurement of a weak incoherent light source (a halogen lamp) is performed. As a stable classical light source, its second order correlation is expected to be $\Gt(\tau)=const., \forall \tau$. Due to the CT, $\Gt(0)>\Gt(\infty)$, and this gap is used to quantify $P^{CT}_n$ as follows:
\begin{equation}
    P^{CT}_n(x,y) = \frac{\Gt_n(x,y,0)-<\Gt_n(x,y,\infty)>}{I_0^T(x,y) + I^T_n(x,y)},
    \label{eq:P_CT}
\end{equation}
where 
\begin{equation}
    I^T(x,y) = \sum_{t=0}^{T} I(x,y,t),
    \label{eq:intImg}
\end{equation}
and the rest of the notations are as in the previous equations.

\begin{figure}[htbp]
    \centering
    \begin{subfigure}[t]{0.45\textwidth} 
        \centering
        \includegraphics[width=\linewidth]{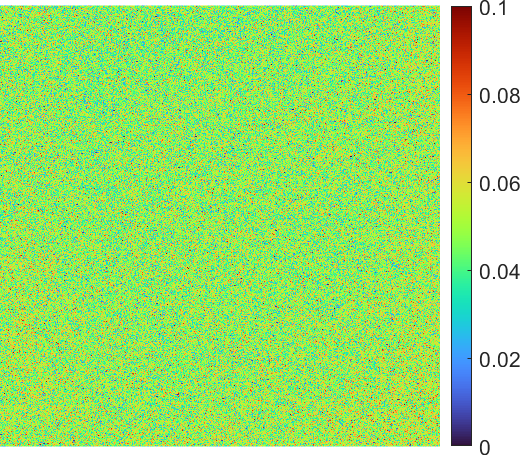}
        \caption{$p^{CT}$ map [\%]}
        \label{fig:p_CT_map}
    \end{subfigure}
    \hfill
    \begin{subfigure}[t]{0.38\textwidth} 
        \centering
        \includegraphics[width=\linewidth]{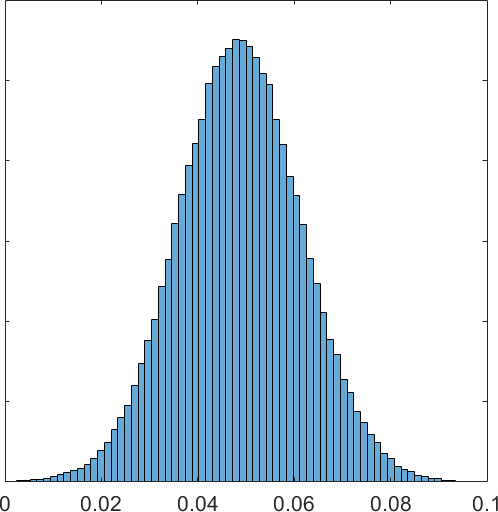}
        \caption{$p^{CT}$ histogram [\%]}
        \label{fig:p_CT_hist}
    \end{subfigure}
    
    \caption{{\bf{Crosstalk characterization of the SS3 detector.}} (a) Map of the estimated crosstalk probability (in percent) for the right neighbor at every pixel. (b) Histogram of the right neighbor crosstalk probability (in percent) across all pixels. The mean crosstalk probability is $\mu=0.05\% $ and the standard deviation is $\sigma=0.02\%$.}
    \label{fig:p_CT}
\end{figure}

As can be seen in \cref{fig:p_CT} (for the right neighbor for example), $P^{CT}_n$ has a random pattern with some spatial variance, which is quite significant as shown in the histogram (\cref{fig:p_CT_hist}). Consideration of this variance turns out to be critical for the proposed scheme. This map (and the similar maps for the other neighbors) is used for the CT compensation, as presented in \cref{sec:02_sub_pipeline}. 

After compensating for CT, the next prominent issue is DC. A DC can generate a false pair by coinciding with a true detection or with another DC, in every delay under consideration. The probability for a DC detection $P^{DC}(x,y)$ is estimated using a long measurement of a closed cap sensor (see the appendix for details), and the median $P^{DC}\approx0.015\%$, with some spatial variance. Using this mapping, DC false pairs are compensated for statistically, as detailed in \cref{sec:02_sub_pipeline}.

The last error factor to be handled is hot pixels, which are pixels with very high DC rate (chosen as the top 3\%). These pixels and their four edge-sharing neighbors are excluded from the correlation measurement. In addition, in the case of extremely hot pixels ($P^{DC}>95\%$), both their edge-sharing, corner sharing, and second-degree neighbors are omitted, due to high values of CT around them. In total, $\mathord{\sim} 15\%$ of the pixels are omitted. However, the sensitivity to hot pixels is expected to be considerably improved with higher SNR (achieved using time-gating and a micro-lens array).

\subsection{Data processing pipeline} \label{sec:02_sub_pipeline}
As discussed in the previous subsection, direct calculation of $\Gt_n(x,y,\tau)$ (the photon correlations using the measured frames) will lead to results that include false pairs originated in both CT and DC. Therefore, these false pairs are compensated for statistically. 

The statistical estimation for the amount of CT false pairs is:
\begin{equation}
    CT_n(x,y) = P^{CT}_n(x,y)\cdot \left( I_0^T(x,y) + I^T_n(x,y)\right).
\end{equation}

Subsequently, the DC false pairs estimation is: 
\begin{equation}
    DC_n(x,y) = I_0^T(x,y)\cdot P^{DC}_n(x,y) + I^T_n(x,y)\cdot P^{DC}(x,y) - P^{DC}(x,y)\cdot P^{DC}_n(x,y)\cdot T,
\end{equation}
where the first and second terms represent false pairs of intensity count with a DC. Since $I^T$ includes both signal and DC components, each term accounts for a pair generated by a countable event (either a real photon or a DC) paired with a DC, in both directions. However, because DC/DC false pairs are included in both the first and second terms, the third term adjusts for this double-counting.

Both estimations are then used to compensate the raw correlation measurement:
\begin{equation}
    \Gt_{n|corr}(x,y,\tau) = 
        \begin{cases}
      \Gt_n(x,y,\tau) - CT_n(x,y) - DC_n(x,y)& \text{$\tau=0$}\\
      \Gt_n(x,y,\tau) - DC_n(x,y) & \text{$\tau\neq 0$}\\
    \end{cases}. 
    \label{eq:G2_corr}
\end{equation}

After achieving a corrected photon correlation map for each neighboring pixels pair, aggregation of the signal from the relevant neighbor-pairs (four edge-sharing neighbors in the current case) is performed by summing the $\Gt_{n|corr}(x,y,\tau)$ functions: 
\begin{equation}
    \Gt_{tot}(x,y,\tau) =  \sum_{n=1}^{N} \Gt_{n|corr}(x,y,\tau).
\end{equation}

Finally, the $\Gt_{tot}(x,y,\tau)$ function is normalized to get $\gt_{tot}(x,y,\tau)$ by:
\begin{equation}
    \gt_{tot}(x,y,\tau) =  \frac{\Gt_{tot}(x,y,\tau)}{<\Gt_{tot}(x,y,\infty)>},
\end{equation}

where
\begin{equation}
    <\Gt(x,y,\infty)> = \frac{1}{\tau_{max}-2} \sum_{\tau=2}^{\tau_{max}} \Gt(x,y,\tau).
    \label{eq:Gt_inf}
\end{equation}

Note that approximating $\Gt(x,y,\infty)$ to be an average of the $\tau=[2,\tau_{max}]$ delays is an arbitrary choice (depends mainly on the SNR).

\section{Experimental Results}
\label{sec:03_exp}
\begin{figure}[htbp]
    \centering
    \begin{subfigure}{0.4\textwidth}
        \centering
        \includegraphics[width=\linewidth]{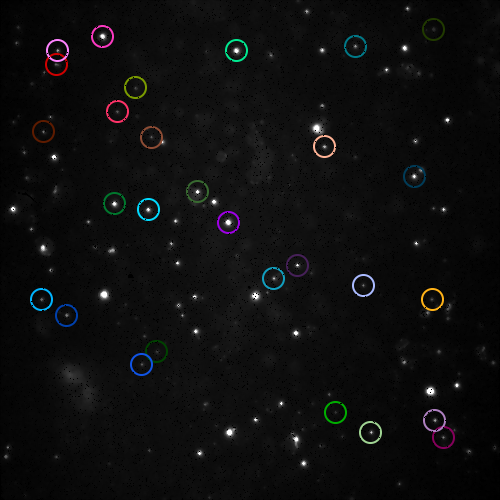} 
        \caption{Intensity image} 
        \label{fig:exp_int}
    \end{subfigure}
    \hfill 
    \begin{subfigure}{0.43\textwidth}
        \centering
        \includegraphics[width=\linewidth]{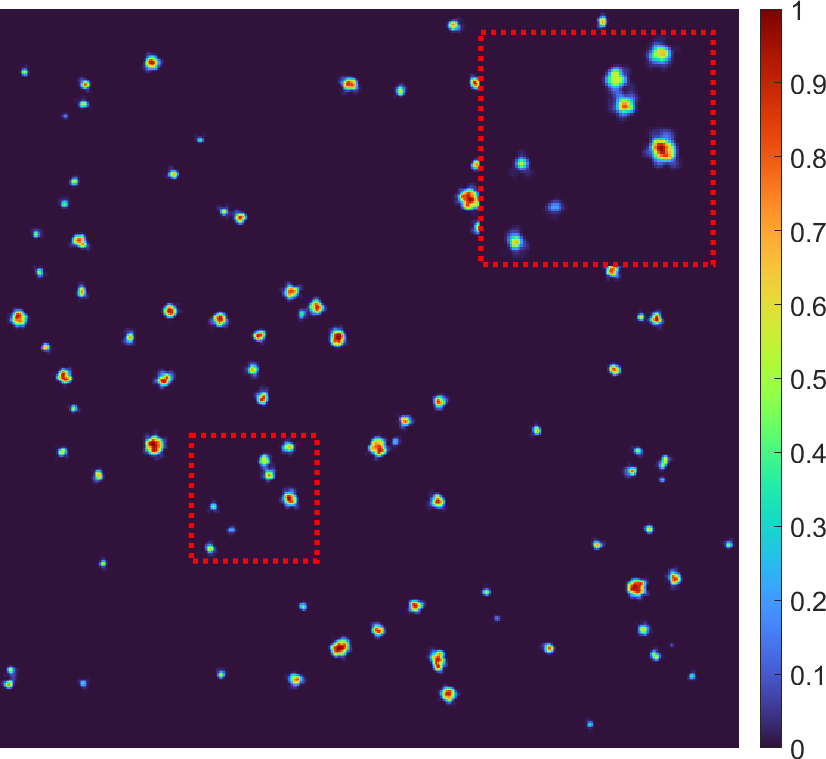} 
        \caption{Wide-field $\gt(0)$ image} 
        \label{fig:WF_g2}
    \end{subfigure}
    
    \vspace{0.3cm} 
    
    \begin{subfigure}{0.4\textwidth}
        \centering
        \includegraphics[width=\linewidth]{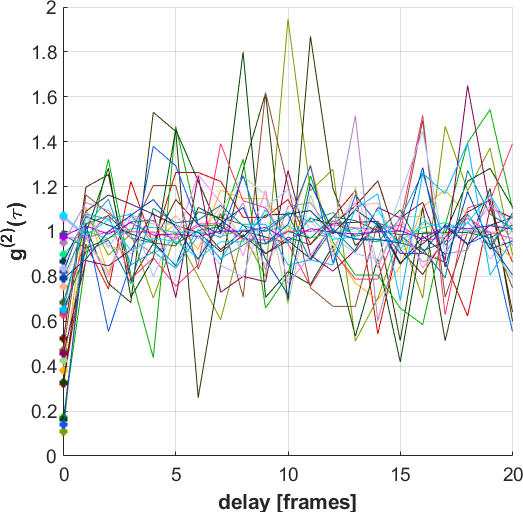} 
        \caption{$g^{(2)}(\tau)$ of the marked points} 
        \label{fig:exp_g2}
    \end{subfigure}
    \hfill 
    \begin{subfigure}{0.45\textwidth}
        \centering
        \includegraphics[width=\linewidth]{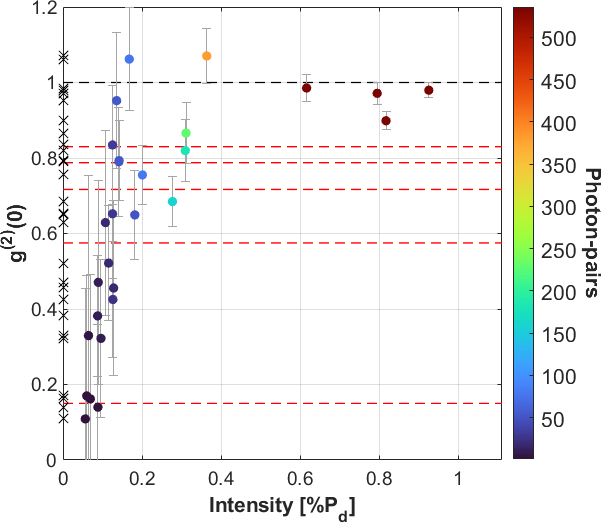} 
        \caption{$g^{(2)}(0)$ vs. Intensity scatter plot} 
        \label{fig:exp_g2_sctr}
    \end{subfigure}
    
    \caption{{\bf{Intensity map and photon statistics at selected locations.}} (a) Integrated intensity image across the entire detector.  (b) Wide-field $\gt(0)$ image (binary mask is used to suppress background noise). Inset on the top-right is a zoom-in on the dashed-red rectangle; note the various $\gt(0)$ values observed. (c) $g^{(2)}(\tau)$ traces for representative points marked by colored circles in the intensity map. (d) $g^{(2)}(0)$ vs. Intensity scatter plot. The color of each dot represents the number of collected photon pairs, according to the colorbar on the right. error-bars indicate the standard deviation of the $\gt(\infty)$ estimation. Dashed lines indicate expected $g^{(2)}(0)$ values. `X' marks on the Y-axis indicate the marginal $\gt(0)$ distribution; low values seem clustered, indicating the number of emitters.}
    \label{fig:exp_int_and_g2}
\end{figure}

Using the setup and data processing procedure described in \Cref{sec:02_method}, several samples of spin coated QD solution (Invitrogen Qdot 545 ITK) were measured. Example results of a $\mathord{\sim}120s$ ($12\cdot10^6$ frames) measurement are presented in \cref{fig:exp_int_and_g2}. In the current setting (without a micro-lens array), the empirical photon detection probability per QD per frame is $P_d\approx0.1\%$. This value aligns with theoretical expectation, as the quantum-yield of the QDs is $\mathord{\sim}50\%$, the objective collects about one third of the emitted photons, the sensor's fill-factor is $\mathord{\sim}10\%$, the per-pixel efficiency is $\mathord{\sim}50\%$, and each spot mapped to $\mathord{\sim}3\times3$ pixels. This stresses the importance in the CT and DC calibration, as $P^{CT}\approx 0.5P_d$, and $P^{DC}\approx 0.15P_d$ (note that the comparison between $P^{CT}$ and $P_d$ is not a direct comparison between two independent constants, as a CT event depends on a countable detection).

The second order correlation function $g^{(2)}(\tau)$ is calculated for $\tau=[0,20]$ (where the inherent temporal resolution is the frame time).  For reference, the same QDs have been measured in a standard HBT setup (see the appendix for details), and the single emitter anti-bunching is estimated around $\gt_{sngl}(0)\approx 0.15$.
To visualize the potential in such wide-field $\gt$ measurement, the intensity and $\gt(0)$ images are presented in \cref{fig:exp_int,fig:WF_g2}. To avoid hot-pixels related artifacts (in both the hot-pixels and their immediate neighbors), their values are replaced with inward interpolation. In addition, to suppress background noise in regions where $\gt(0)$ cannot be reliably estimated due to low signal, the raw $\gt(0)$ map is multiplied with a binary mask (with threshold of 3 times the mean value) that filters out the background 
 and keeps the values around the peak intensity points. Qualitatively, the $\gt(0)$ values are correlated with the brightness and size of the intensity values; this hints the ability to utilize the $\gt(0)$ values to estimate the number of emitters in every spot.
To further check this quantitatively, $\gt(\tau)$ traces for several representing points are presented in \cref{fig:exp_g2} (to avoid hot-pixels related issues, only points without neighboring hot-pixels were post selected). As expected, for visibly large aggregates that contain a large number of QDs (each an independent single-photon emitter), statistics approach those of classical uncorrelated light, where $g^{(2)}(0)\approx 1$. However, for the smaller, diffraction-limited points, that might contain a single or few QDs, $\gt(0)$ is significantly below unity, as expected ($\gt(0)=1-\frac{1-\gt_{sngl}(0)}{n}$ for $n$ identical single-photon emitters). As can be seen, the reference $\gt_{sngl}(0)$ correspond with the lowest values observed in the wide-field measurement.

More insight can be obtained by plotting $g^{(2)}(0)$ vs. intensity values at the same pixels (\cref{fig:exp_g2_sctr}). A correlation is observed between low intensities and low $\gt(0)$ values, both indicating fewer QDs within the observed spot. High intensity points ($P_d > 0.6$) can be clearly labeled as containing multiple QDs by both intensity and photon correlation. However, at lower intensities, the $\gt(0)$ values uncover new information, by indicating points with similar intensity containing different number of emitters. However, before utilizing this measurement for emitter counting, a careful error analysis is required. We note that while the error bars in \cref{fig:exp_g2_sctr} (which indicate $\Delta \gt(0)\approx\frac{\Delta \Gt(0)}{\Gt(\infty)}$, approximated by the standard deviation of $\gt(\infty)$) for the lower intensity data points are significant, they already allow a reliable separation between single emitters and multiple emitters, and also some discrimination ($\pm1$) up to $n=3$. The hardware limitations giving rise to the large error bars are expected to be alleviated in the future, as discussed in detail in the following section, enabling an even larger dynamical range for emitter counting. Additional experimental results are provided in the appendix.

\begin{figure}[htbp]
\centering\includegraphics[width=7cm]{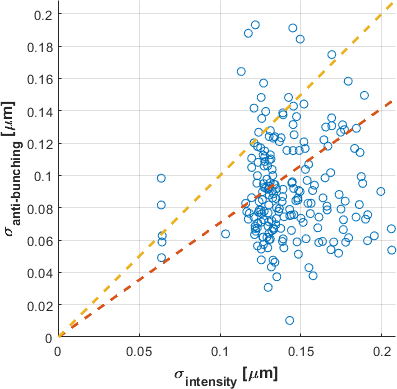}
\caption{{\bf{SR modality resolution analysis.}} To quantify the resolution enhancement in the proposed anti-bunching contrast image, Gaussian function are fitted to corresponding coordinates (of intensity peaks) in both the intensity and anti-bunching images, and a scatter plot of the Gaussian $\sigma$ values is presented (the scatter plot contains analysis from three different measurements). While the spread of values originate in the anti-bunching is quite large (due to its inherent noise and the low SNR of the current setting), the values are scattered around the expected $\sqrt2$ line (dashed red). In addition, the vast majority of the points are below the unity line (dashed yellow), which indicates resolution improvement.}
\label{fig:SR_scatter}
\end{figure}

One potential application of such information is SR microscopy, using the anti-bunching image (i.e. image of the `missing pairs' at the zero delay- $\Gt(x,y,\infty)-\Gt(x,y,0)$) as the contrast, as presented in \cite{schwartz2012improved, schwartz2013superresolution}. While intensity images are clearly less noisy, the anti-bunching images provide higher resolution and eliminate background fluorescence (similar to classical fluctuation-based correlation microscopy). Methods to fuse such images to a joint clean SR image, have recently been proposed \cite{Uri_jointSR,Lior_jointSR}. In the current experimental setting, the resulting anti-bunching images are quite noisy due to the low SNR, therefore the resolution enhancement is not easily visible (see appendix). Therefore, to estimate the resolution improvement statistically, three different measurements are taken (the one presented in \cref{fig:exp_int_and_g2} and two similar measurements, presented in the appendix). Gaussian functions are fitted to corresponding local maxima locations in the intensity and anti-bunching images, and the standard deviations of each Gaussian pair are compared in a scatter plot (\cref{fig:SR_scatter}). Due to the high noise level in the anti-bunching images, the spread of values is large, however it is centered around the expected theoretical value of $\sqrt{2}$ resolution improvement (as presented in \cite{schwartz2012improved, schwartz2013superresolution}). 

\section{Discussion}
\label{sec:04_discussion}
As presented in \Cref{sec:03_exp}, the proposed method enables wide-field imaging while obtaining per-pixel photon statistics information. As we have demonstrated, this information by itself can be directly used to perform SR imaging. Moreover, this scheme holds potential for various advanced microscopy applications. For example, using the relation $g^{(2)}(0)=1-\frac{1-\gt_{sngl}(0)}{n}$, the value of the second order correlation can be used to estimate the number of unresolved emitters generating a diffraction limited spot. This insight can be utilized as valuable side-information for single molecule localization microscopy (SMLM) based SR imaging methods. For example, in STORM \cite{STORM} and PALM \cite{PALM}, an indication of the number of emitters can help differentiate if a spot includes a single emitter or more than one; this additional information can be used to considerably speed up the imaging process using such methods, as it may relax the constraints on emitter sparsity in every frame. Similarly, in microscopy methods based on fluorescence fluctuations \cite{SOFI}, an estimation to the number of emitters can be used as a supplement for advanced SR algorithms. We note that improved SNR is required for accurate emitter number estimation (as discussed in the following paragraph). However, this is unnecessary for the SR imaging modality (discussed shortly in \cref{sec:03_exp} and in the appendix in detail), which relies only on the deviation of $\Gt(0)$ from $\Gt(\infty)$, and is thus insensitive to a possible offset for $\Gt(0)$.

As demonstrated in this work, the current implementation's SNR allows wide-field emitter counting up to $n{\sim}3$ in a 2-minute measurement, at $P_d\approx0.1\%$. However, several hardware and firmware enhancements, that have already been implemented in similar sensors, can extend this range significantly. Micro-lens arrays are an established technique to increase the effective fill-factor of SPAD arrays. Micro-lenses on previous generation SwissSPAD were shown to enhance the effective photon detection efficiency by a factor of 4.2 \cite{Bruschini_microlenses}. As the second-order correlation signal scales quadratically with $P_d$, this will enhance the signal-to-noise ratio by more than an order-of-magnitude. Additionally, SS3 supports time gating (not implemented in the current firmware). The fluorescence signal features distinct temporal characteristics (lifetime of $\tau \sim \SI{e-8}{s}$ after excitation), while the DC is time-independent over the $\SI{10}{\mu s}$ acquisition time of the frame. By time-gating, that is keeping detections in the ${\sim}\SI{100}{ns}$ following any excitation pulse and discarding other detections, the DC noise can be decreased by two orders of magnitude (see supporting information of \cite{Gur_heralded}). Together, they can allow for well above an order of magnitude increase in speed. A future implementation of on-chip correlation calculation (which is in active research and development) will enable to dramatically increase the frame-rate, thus resulting in much faster acquisition time. Combining these innovations can lead to an at least order of magnitude faster acquisition of SR microscopy images compared to existing imaging methods.

\section{Summary}
\label{sec:05_discussion}

In this work, we have introduced a method for rapid, massively-multiplexed sensing of photon statistics using a $\SI{0.25}{Mpix}$ single-photon avalanche diode array. Applying this apparatus to investigate a two-dimensional sample of single-photon emitters (quantum dots), we demonstrate utilizing wide-field photon statistics measurements for emitter counting and super resolution imaging, and discuss other sensing and imaging opportunities. Crucially, this was achieved with a technological prototype, the SwissSPAD3. Several readily available hardware and firmware enhancements, such as micro-lens arrays and temporal gating, can support orders-of-magnitude enhancement of signal-to-noise, substantially extending the performance and range of relevant applications.

\begin{backmatter}

\bmsection{Acknowledgments}
Financial support by a grant from KLA-Tencor is gratefully acknowledged. D.O. is the incumbent of the Harry Weinrebe professorial chair of laser physics.

\bmsection{Disclosures}
Edoardo Charbon is co-founder of Novoviz, and Claudio Bruschini and Edoardo Charbon are co-founders of Pi Imaging Technology. Neither company has been involved with the work or paper.

\bmsection{Data availability} Data underlying the results presented in this paper are not publicly available at this time but may be obtained from the authors upon reasonable request.

\end{backmatter}

\section*{Appendix}
\appendix

\section{Dark-counts Characterization}

\begin{figure}[htbp]
    \centering

    \begin{subfigure}[b]{0.3\textwidth} 
        \centering
        \includegraphics[width=\textwidth]{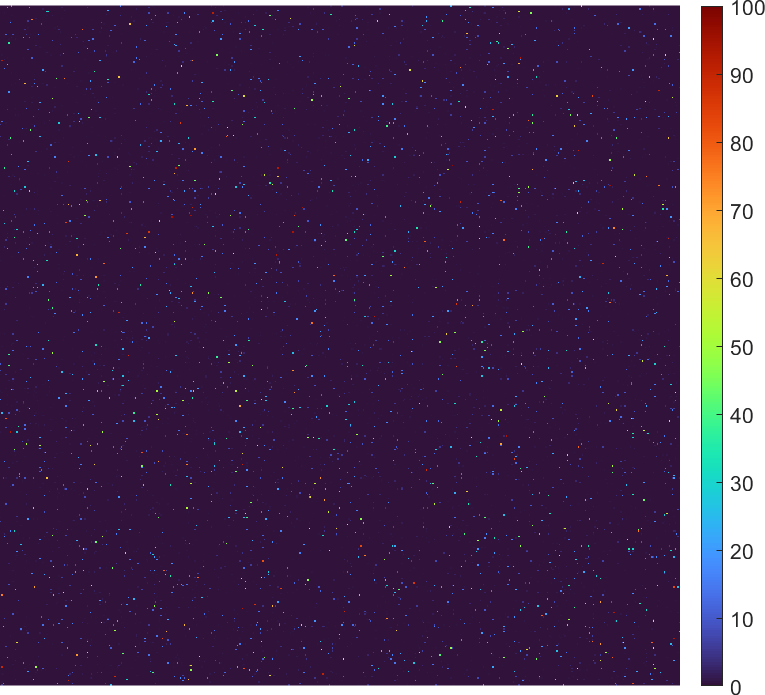} %
        \caption{$P^{DC}(x,y)$ map} 
        \label{fig:dcrMap_full}
    \end{subfigure}
    \hfill
    \begin{subfigure}[b]{0.3\textwidth} 
        \centering
        \includegraphics[width=\textwidth]{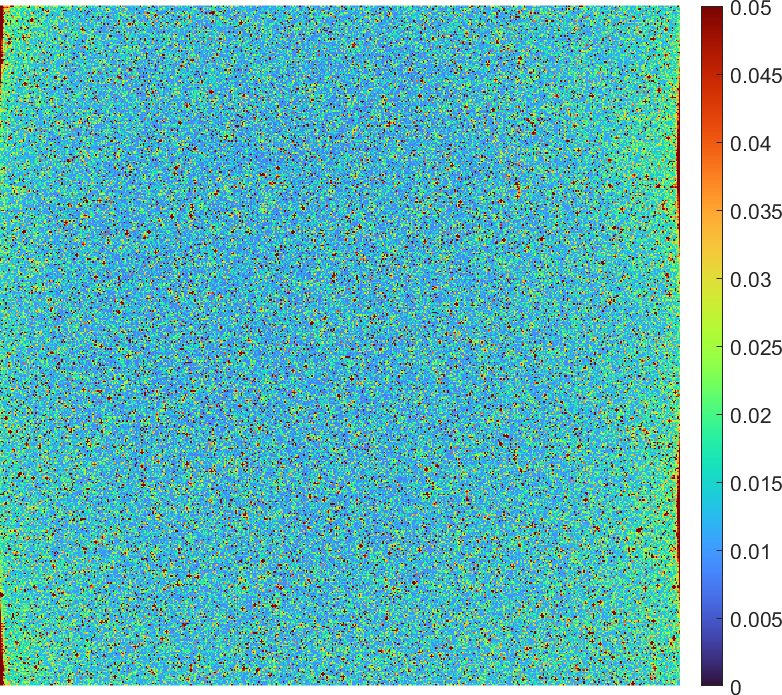} 
        \caption{$P^{DC}(x,y)$ map - stretched} 
        \label{fig:dcrMap_strch}
    \end{subfigure}
    \hfill
    \begin{subfigure}[b]{0.3\textwidth} 
        \centering
        \includegraphics[width=\textwidth]{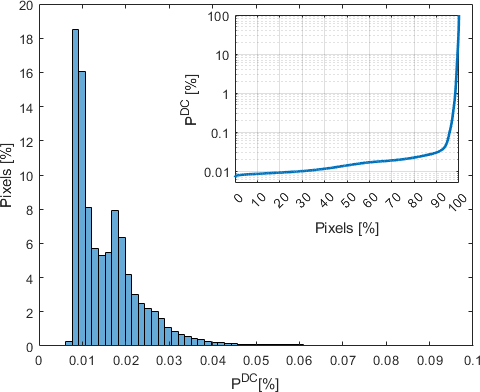} 
        \caption{$P^{DC}(x,y)$ histogram and cumulative plot}
        \label{fig:dcrHist}
    \end{subfigure}
    
    \caption{{\bf{$P^{DC}$ map and histogram.}} To compensate for pairs originated due to DC, the $P^{DC}(x,y)$ map is estimated. (a) $P^{DC}(x,y)$ map (full dynamic range). As can be seen, while almost all of the pixels have very low $P_{DC}$, about $\mathord{\sim}5\%$ of the pixels have very high values comparing to the typical $P_d$. Since each pixel affects its immediate 4 or 8 neighbors, a careful compensation of this error is required in order to keep the effective area as large as possible. (b) $P^{DC}(x,y)$ map (stretched dynamic range to $[0,0.05\%]$). Focusing on the lower values, the spatial distribution of the DC rate can be observed. (c) Histogram of $P^{DC}(x,y)$  unveils the distribution of the 95\% of pixels with low $P_{DC}$ values. As can be seen, the histogram comprises 'cold' and 'warm' peaks, with a very long tail of 'hot pixels' (almost reaching $P_{DC}=100\%$ for very few pixels). }
    \label{fig:dcrMapAndHist}
\end{figure}

As discussed in Section 2 of the paper, in order to calibrate the photon pairs detected due to dark-counts (DC), a map of the pixel-wise probability of a DC, $P^{DC}(x,y)$, is required. To obtain such a map, a $\mathord{\sim}160s$ measurement of a covered sensor was performed. By averaging the binary frames, $P^{DC}$ map per frame is estimated. The map and its histogram are presented in \cref{fig:dcrMapAndHist}. As can be seen, some spatial variance exists in the $P^{DC}$ map (see \cref{fig:dcrMap_strch}), and its proper compensation is critical for the proposed task of wide-field photon correlation sensing. As can be seen in the histogram, pixels can be grouped to three: most of the sensor contains 'cold' pixels (left peak in the histogram), with a relatively low $P_{DC}$ value; the second population is the 'warm' pixels (right peak in the histogram), with a bit higher $P_{DC}$, but that still can be compensated well enough; the last group (about 5\% of the pixels) comprises the `hot' pixels (the long tail in the histogram, partially visible in \cref{fig:dcrHist}). As such `hot' pixels affect both them and their neighbors, if not compensated well they have to be excluded from the measurement. Therefore, the effort to compensate the DC-related pairs for large $P_{DC}$ values directly affects the effective area of the sensor. In the current work, the top 3\% of hot pixels are excluded from the measurement; together with their 4-immediate neighbors, about 15\% of the sensor area is effectively excluded.

We note that the sensitivity to DC is related to the detection efficiency (or probability of detection) $P_d$. As discussed in Section 4 of the paper, the specific sensor being used in this work does not comprise a micro-lens array, and time gating is not implemented, which lead to increased sensitivity to DC. As both of these technical limitations are expected to be solved, the sensitivity to DC can be considerably lower, and 'hotter' pixels can be included in the analysis, leading to larger effective area utilization.

\section{Bias due to frequency mismatch}

As discussed in Section 2 of the paper, to achieve a condition of (at most) single-photon per-frame per quantum-dot (QD), two conditions should be met: (1) the illumination pulse-width should be considerably shorter than the QDs radiative lifetime, and (2) a single illumination pulse per frame is required. The pulse width requirement is easily met, as the pulse width is $t_p=300fs$, while the QDs radiative lifetime is $\tau_{QD}>50ns$. To address the single-pulse per-frame requirement, ideally, the laser repetition rate should be synchronized to the frame acquisition rate, or set asynchronously to $f_{LSR}=f_{SS3}-\epsilon$ (the SS3 native frame-rate is $f_{SS3}=97.7 kframes/s$.). Due to technical firmware limitations, the laser is not synchronized to the SS3, and its repetition rate is marginally higher ($f_{LSR}=100 kHz$), which results in $2.3\%$ of the frames that are illuminated with two pulses. This issues results in some bias in the $\gt$ measurement, as in such double-illuminated frames a false-pair can be detected. Such false-pair event occurs when the first pulse generates a detectable photon in a specific pixel, and the second pulse generates a detectable photon in one of its 4 or 8 neighbors (depends on which neighbors are taken under consideration). Since the QDs quantum yield is $\Phi\approx50\%$ and the overall detection efficiency (per QD per pixel per frame) is very low ($P_d\approx0.1\%$), the bias due to the possibility to detect such false pair is very low (for example, for a $12\cdot10^6$ frames measurement like the one presented in the paper, it results in $\mathord{\sim}0.25$ pairs bias). Comparing to the present errors (estimated in the errorbars in Fig.~4 of the paper), this bias is negligible.

\section{Reference anti-bunching measurement}

\begin{figure}[htbp]
\centering\includegraphics[width=7cm]{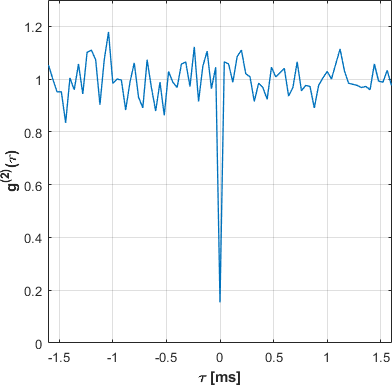}
\caption{{\bf{Reference $\gt(\tau)$ measurement.}} Using a standard HBT setup, the same QDs were measured to achieve a reference $\gt(\tau)$ estimation. As can be seen, $\gt(0)\approx0.15$, which is in good correspondence with the lowest values measured in the wide-field setup.}
\label{fig:g2_HBT}
\end{figure}

To compare the $\gt$ values measured in the wide-field setup to an external independent reference measurement, QDs from the same batch were measured in a standard HBT experimental setup. The setup is based on a commercial inverted microscope (Eclipse Ti-U, Nikon) and the sample is illuminated using a pulsed diode laser with a wavelength of $470 nm$ and a repetition rate of $5 MHz$ (LDH-P-C-470B, PicoQuant). The laser beam is focused on the sample plane using a high numerical-aperture (NA) oil immersion objective lens ($\times100$, $1.3 NA$, Nikon) which also collects the resulting fluorescence light. Back-scattered laser light is filtered out by a dichroic mirror (505 LP, Chroma) and a long pass dielectric filter (488 LP, Semrock). The fluorescence signal, collected from a single QD located in the illumination spot, is split to two commercial SPAD detectors (COUNT-20B, Laser Components) using a 50/50 fiber beam-splitter. The $\gt(\tau)$ results are presented in \cref{fig:g2_HBT}. As can be seen, the anti-bunching value $\gt(0)\approx0.15$, which is in good correspondence with lower values measured using the wide-field setup, as presented in Figure 4 of the paper. 

\section{Resolution enhancement analysis}

\begin{figure}[htbp]
    \centering
    \begin{subfigure}[b]{0.32\textwidth}
        \includegraphics[width=\textwidth]{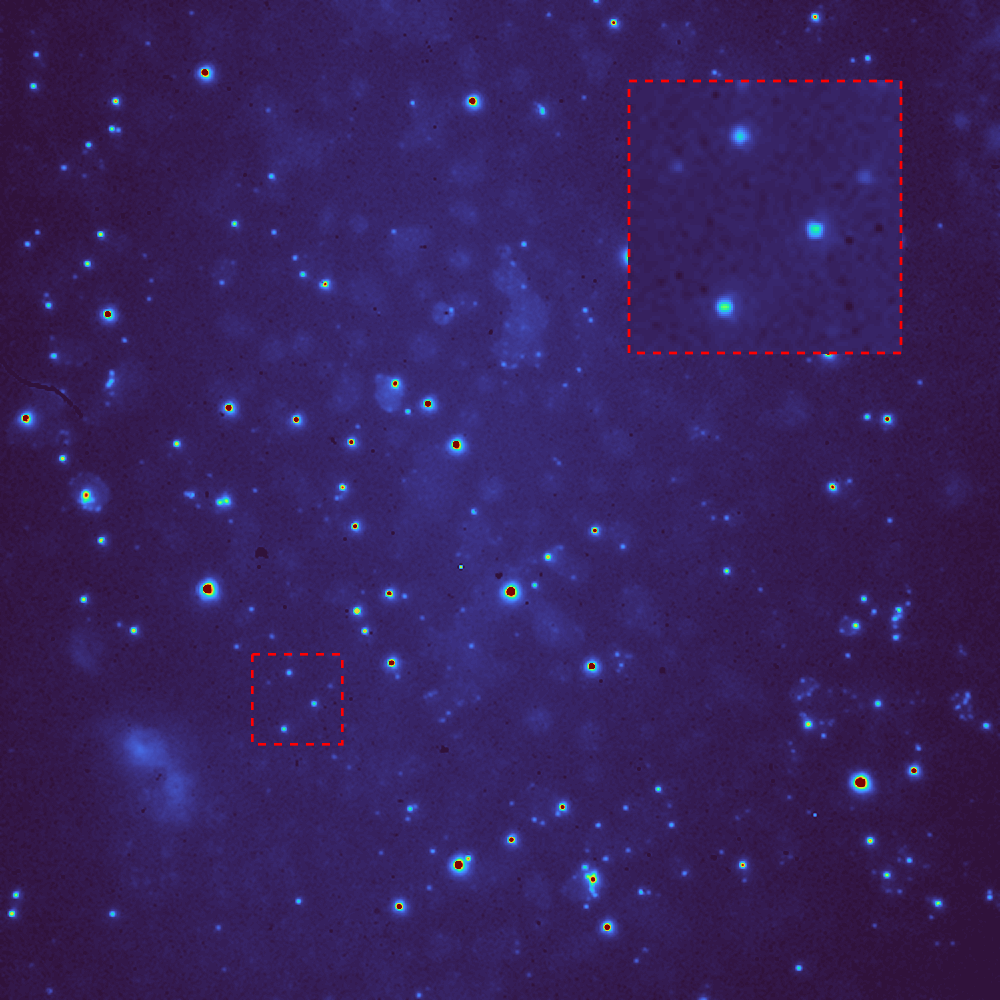}
        \caption{Intensity image}
        \label{fig:SR_int}
    \end{subfigure}
    \hfill
    \begin{subfigure}[b]{0.32\textwidth}
        \includegraphics[width=\textwidth]{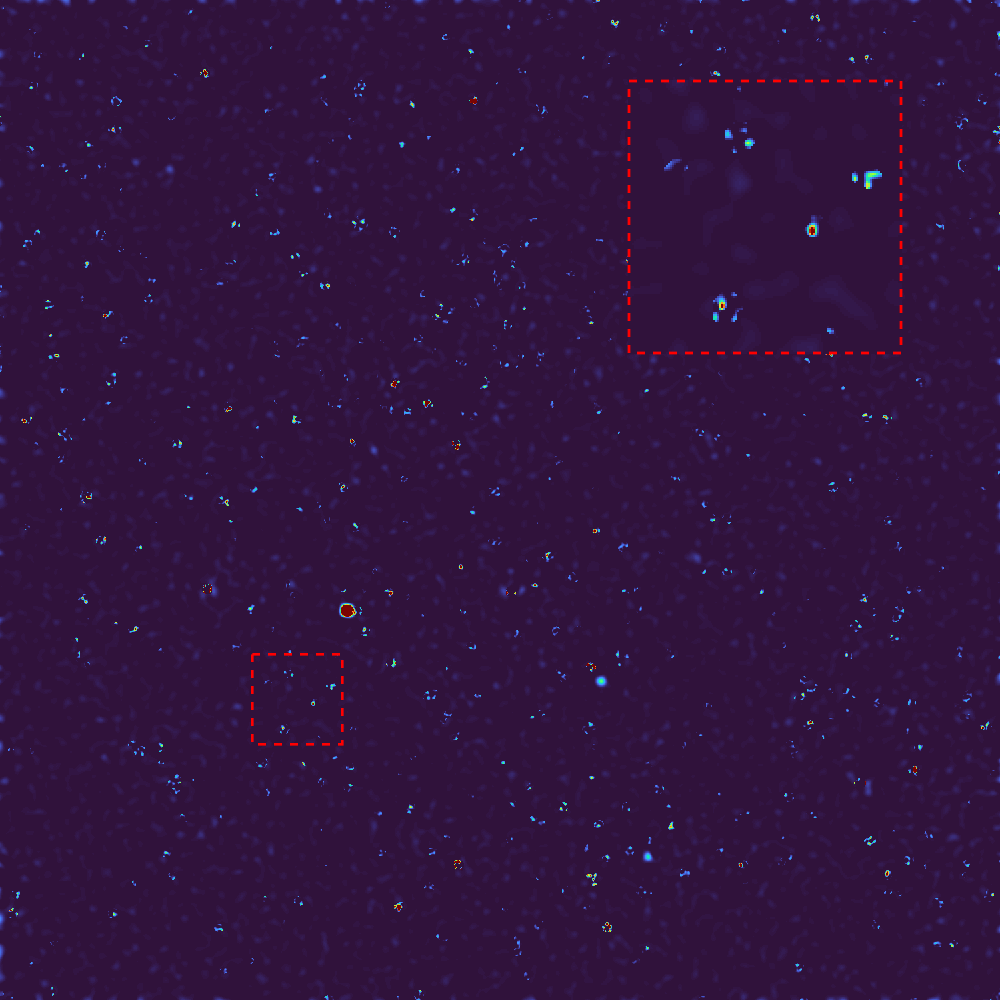}
        \caption{Anti-bunching image}
        \label{fig:SR_abch}
    \end{subfigure}
    \hfill
    \begin{subfigure}[b]{0.32\textwidth}
        \includegraphics[width=\textwidth]{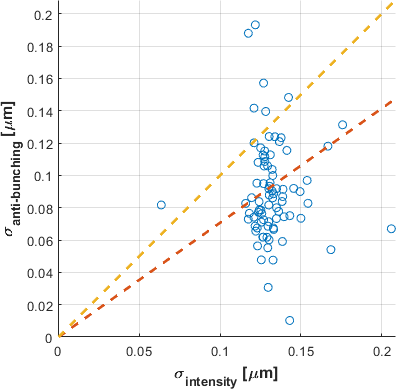}
        \caption{Gaussian fitting $\sigma$ scatter plot}
        \label{fig:SR_sctr}
    \end{subfigure}
    \caption{{\bf{Resolution enhancement analysis.}} Using anti-bunching as a the contrast, a resolution enhancement of $\sqrt2$ is expected. Comparing the (a) intensity image and (b) anti-bunching image, this enhancement can be observed (note that both images are processed for display purposes). To quantify this enhancement, Gaussian function were fitted to intensity peaks, and a (c) scatter plot of corresponding Gaussian $\sigma$ values is presented. While the spread of values originate in the anti-bunching is quite large (due to its inherent noise and the low SNR of the current setting), the values are scattered around the expected $\sqrt2$ line (plotted in red).}
    \label{fig:SR}
\end{figure}

As discussed in Section 3 of the paper, a direct potential application of wide-field $\gt$ measurements is super-resolution (SR) microscopy. By taking the anti-bunching image (i.e. image of the 'missing pairs' at the zero delay- $\Gt(x,y,\infty)-\Gt(x,y,0)$) as the contrast, a resolution enhancement of $\sqrt2$ is expected, as presented in \cite{schwartz2012improved, schwartz2013superresolution}. While this enhancement is theoretically straight-forward, it practically requires some oversampling and/or high enough signal-to-noise (SNR) to achieve it. In the current setup both sampling and SNR are limited (due to similar reasons), and in addition $\mathord{\sim}15\%$ of the pixels are excluded from the analysis (`hot' pixels and their immediate neighbors). Since this issue limits the visual qualitative assessment of the resolution enhancement, a statistical analysis is used to estimate it quantitatively. Various local maxima locations (of the intensity image) are taken, and Gaussian functions are fitted around these locations in both the intensity and anti-bunching images. To assess the resolution improvement, the standard deviations of each Gaussian pair are compared. A scatter plot of the standard deviation ($\sigma$) is presented in \cref{fig:SR} (Note that for display purposes, denoising, hole filling and gamma correction were applied, while the Gaussian fitting had been performed on the raw images). As can be seen, the $\sigma$ of the Gaussians fitted in the intensity image are in good correspondence with the theory (standard deviation of the Gaussian fitted on a diffraction limited spot $\sigma=0.45\frac{\lambda}{NA}$). However, the corresponding fitted values from the anti-bunching image are quite spread, mainly due to the high noise level of the $\gt$ estimation. Despite that, they are centered around the expected theoretical value of $\sqrt{2}$ resolution improvement. As discussed in Section 4 of the paper, the detection efficiency boost from a micro-lens array and the lower sensitivity to DC thanks to gating are expected to provide much cleaner $\gt$ estimation. This will directly lead to much better results in the current configuration, and can also enable denser sampling of each diffraction-limited spot (trade-offing the field-of-view), and by that to better condition the photon-pairs collection (as its probability is $P_{pair}=P_{1}P_{2}$ which is maximized in $P_1=P_2$).

\section{Additional experimental results}

\begin{figure}[htbp]
    \centering
    \begin{subfigure}{0.4\textwidth}
        \centering
        \includegraphics[width=\linewidth]{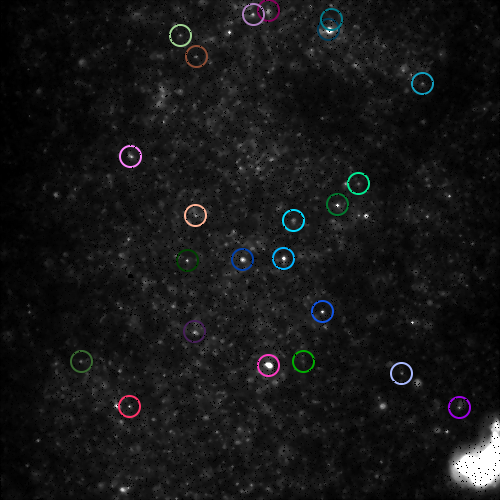} 
        \caption{Intensity image} 
        \label{fig:exp_2_int}
    \end{subfigure}
    \hfill 
    \begin{subfigure}{0.43\textwidth}
        \centering
        \includegraphics[width=\linewidth]{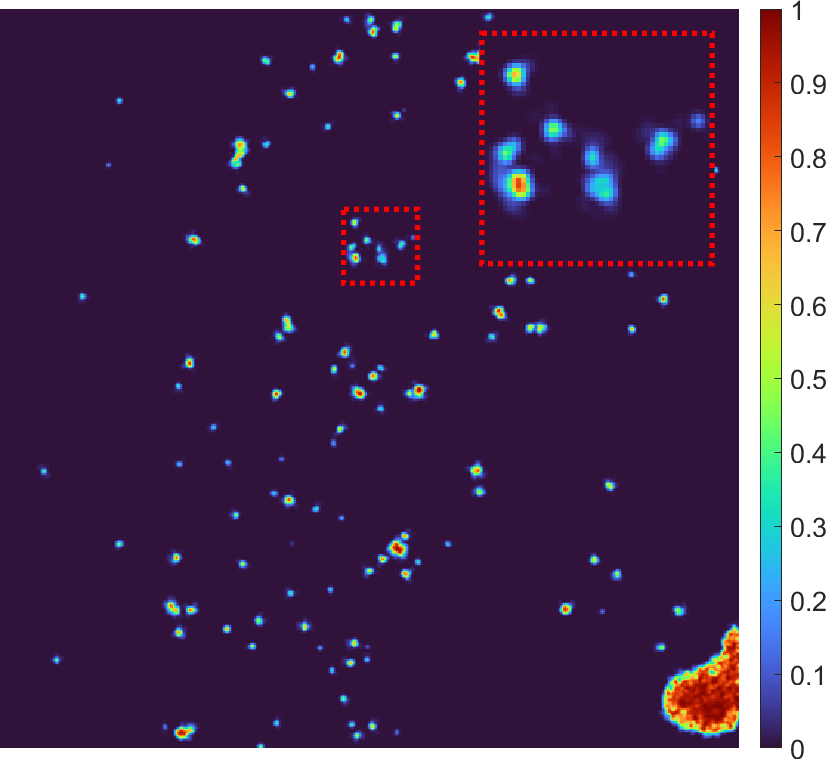} 
        \caption{Wide-field $\gt(0)$ image} 
        \label{fig:exp_2_WF_g2}
    \end{subfigure}
    
    \vspace{0.3cm} 
    
    \begin{subfigure}{0.4\textwidth}
        \centering
        \includegraphics[width=\linewidth]{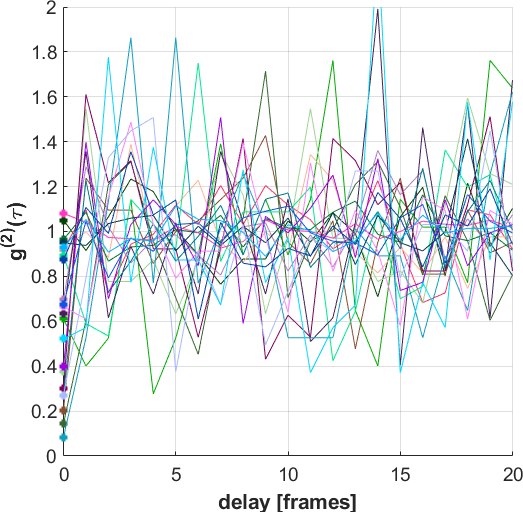} 
        \caption{$g^{(2)}(\tau)$ of the marked points} 
        \label{fig:exp_2_g2}
    \end{subfigure}
    \hfill 
    \begin{subfigure}{0.45\textwidth}
        \centering
        \includegraphics[width=\linewidth]{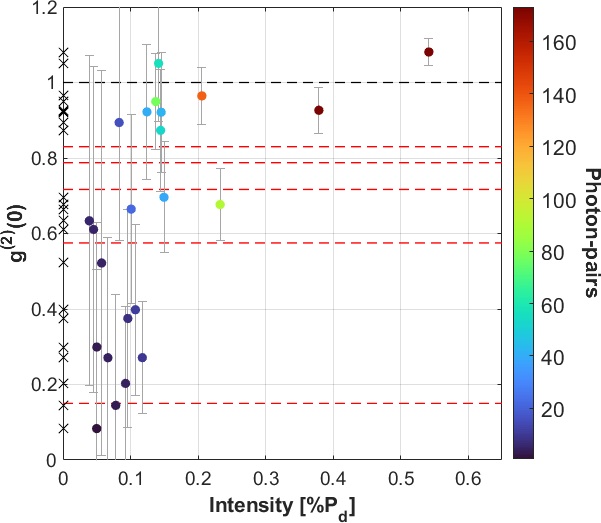} 
        \caption{$g^{(2)}(0)$ vs. Intensity scatter plot} 
        \label{fig:exp_2_g2_sctr}
    \end{subfigure}
    
    \caption{{\bf{Intensity map and photon statistics at selected locations.}} (a) Integrated intensity image across the entire detector. (b) Wide-field $\gt(0)$ image (binary mask is used to suppress background noise). Inset on the top-right is a zoom-in on the dashed-red rectangle; note the various $\gt(0)$ values observed. (c) $g^{(2)}(\tau)$ traces for representative points marked by colored circles in the intensity map. (d) $g^{(2)}(0)$ vs. Intensity scatter plot. The color of each dot represents the number of collected photon pairs, according to the colorbar on the right. Dashed lines indicate ideal $g^{(2)}(0)$ values. }
    \label{fig:exp_2_int_and_g2}
\end{figure}

\begin{figure}[htbp]
    \centering
    \begin{subfigure}{0.4\textwidth}
        \centering
        \includegraphics[width=\linewidth]{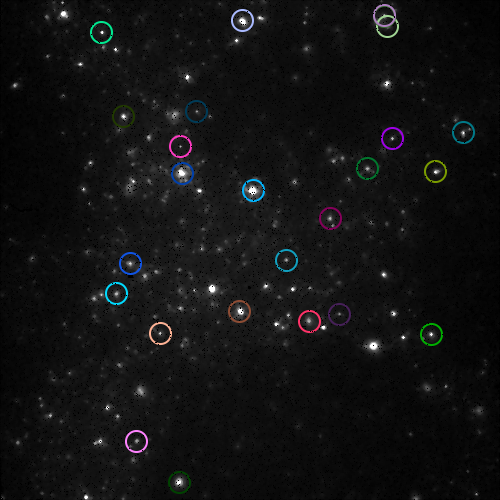} 
        \caption{Intensity image} 
        \label{fig:exp_3_int}
    \end{subfigure}
    \hfill 
    \begin{subfigure}{0.43\textwidth}
        \centering
        \includegraphics[width=\linewidth]{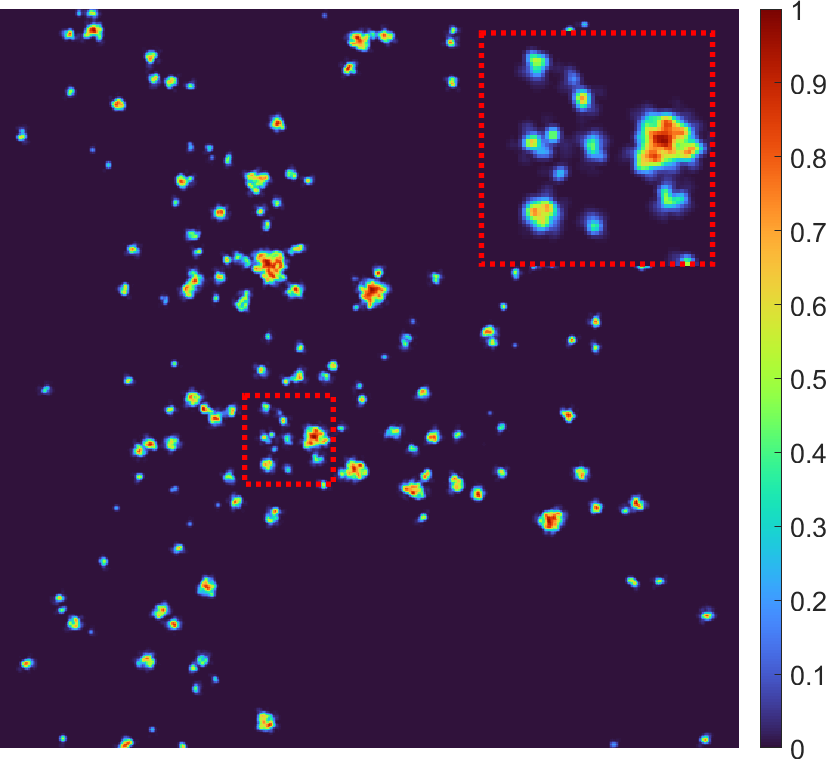} 
        \caption{Wide-field $\gt(0)$ image} 
        \label{fig:exp_3_WF_g2}
    \end{subfigure}
    
    \vspace{0.3cm} 
    
    \begin{subfigure}{0.4\textwidth}
        \centering
        \includegraphics[width=\linewidth]{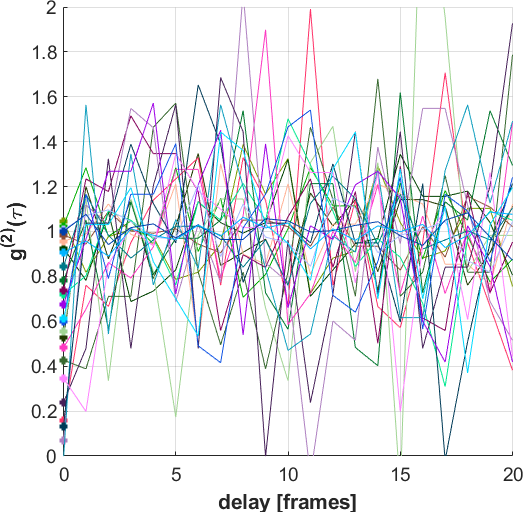} 
        \caption{$g^{(2)}(\tau)$ of the marked points} 
        \label{fig:exp_3_g2}
    \end{subfigure}
    \hfill 
    \begin{subfigure}{0.45\textwidth}
        \centering
        \includegraphics[width=\linewidth]{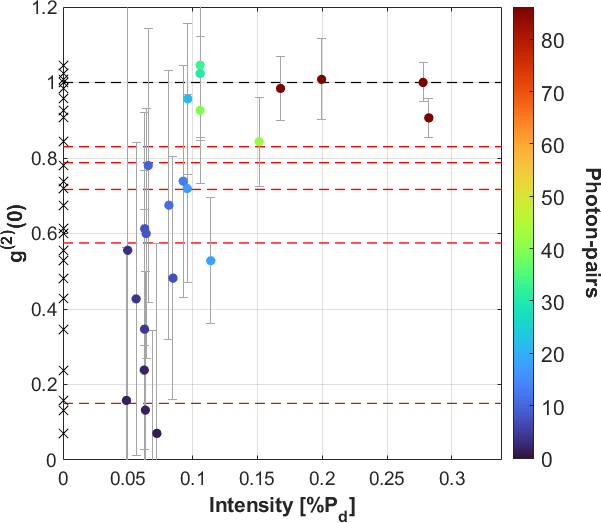} 
        \caption{$g^{(2)}(0)$ vs. Intensity scatter plot} 
        \label{fig:exp_3_g2_sctr}
    \end{subfigure}
    
    \caption{{\bf{Intensity map and photon statistics at selected locations.}} (a) Integrated intensity image across the entire detector. (b) Wide-field $\gt(0)$ image (binary mask is used to suppress background noise). Inset on the top-right is a zoom-in on the dashed-red rectangle; note the various $\gt(0)$ values observed. (c) $g^{(2)}(\tau)$ traces for representative points marked by colored circles in the intensity map. (d) $g^{(2)}(0)$ vs. Intensity scatter plot. The color of each dot represents the number of collected photon pairs, according to the colorbar on the right. Dashed lines indicate ideal $g^{(2)}(0)$ values. }
    \label{fig:exp_3_int_and_g2}
\end{figure}

Following the experimental results presented in Section 3 of the paper, additional results (of other samples from the same QDs batch) are presented (in the same format). A similar length measurement of $\mathord{\sim}120s$ ($12\cdot10^6$ frames) is presented in \cref{fig:exp_2_int_and_g2}, and a bit shorter measurement of $\mathord{\sim}90s$ / $(9\cdot10^6$ frames) is presented in \cref{fig:exp_3_int_and_g2}. As can be seen, although a bit higher noise level on the low estimates, a similar trend in the $\gt$ values is observed.

\bibliography{sample}

\begin{thebibliography}{10}
\newcommand{\enquote}[1]{``#1''}

\bibitem{HBT}
R.~H. Brown and R.~Twiss, \enquote{Lxxiv. a new type of interferometer for use in radio astronomy,} {\protect\JournalTitle{The London, Edinburgh, and Dublin Philosophical Magazine and Journal of Science}} \textbf{45}, 663--682 (1954).

\bibitem{dravins_stellarIntensityRev}
D.~Dravins, \enquote{Intensity interferometry: optical imaging with kilometer baselines,} in \emph{Optical and Infrared Interferometry and Imaging V,}  vol. 9907 (SPIE, 2016), pp. 128--139.

\bibitem{gisin2007quantComm}
N.~Gisin and R.~Thew, \enquote{Quantum communication,} {\protect\JournalTitle{Nature photonics}} \textbf{1}, 165--171 (2007).

\bibitem{couteau2023quanComAndComputing}
C.~Couteau, S.~Barz, T.~Durt, \emph{et~al.}, \enquote{Applications of single photons to quantum communication and computing,} {\protect\JournalTitle{Nature Reviews Physics}} \textbf{5}, 326--338 (2023).

\bibitem{lloyd2008enhanced}
S.~Lloyd, \enquote{Enhanced sensitivity of photodetection via quantum illumination,} {\protect\JournalTitle{Science}} \textbf{321}, 1463--1465 (2008).

\bibitem{england2019quantum}
D.~G. England, B.~Balaji, and B.~J. Sussman, \enquote{Quantum-enhanced standoff detection using correlated photon pairs,} {\protect\JournalTitle{Physical Review A}} \textbf{99}, 023828 (2019).

\bibitem{schwartz2012improved}
O.~Schwartz and D.~Oron, \enquote{Improved resolution in fluorescence microscopy using quantum correlations,} {\protect\JournalTitle{Phys. Rev. A}} \textbf{85}, 033812 (2012).

\bibitem{schwartz2013superresolution}
O.~Schwartz, J.~M. Levitt, R.~Tenne, \emph{et~al.}, \enquote{Superresolution microscopy with quantum emitters,} {\protect\JournalTitle{Nano letters}} \textbf{13}, 5832--5836 (2013).

\bibitem{tenne2019super}
R.~Tenne, U.~Rossman, B.~Rephael, \emph{et~al.}, \enquote{Super-resolution enhancement by quantum image scanning microscopy,} {\protect\JournalTitle{Nature Photonics}} \textbf{13}, 116--122 (2019).

\bibitem{Charbon2013spadBookChapter}
E.~Charbon, M.~Fishburn, R.~Walker, \emph{et~al.}, \emph{{SPAD}-Based Sensors} (Springer Berlin Heidelberg, Berlin, Heidelberg, 2013), pp. 11--38.

\bibitem{bruschini2019spadReview}
C.~Bruschini, H.~Homulle, I.~M. Antolovic, \emph{et~al.}, \enquote{Single-photon avalanche diode imagers in biophotonics: review and outlook,} {\protect\JournalTitle{Light: Science \& Applications}} \textbf{8}, 87 (2019).

\bibitem{SS3}
M.~Wayne, A.~Ulku, A.~Ardelean, \emph{et~al.}, \enquote{A 500$\times$ 500 dual-gate {SPAD} imager with 100\% temporal aperture and 1 ns minimum gate length for {FLIM} and phasor imaging applications,} {\protect\JournalTitle{IEEE Transactions on Electron Devices}} \textbf{69}, 2865--2872 (2022).

\bibitem{Glauber_1963_a}
R.~J. Glauber, \enquote{Photon correlations,} {\protect\JournalTitle{Phys. Rev. Lett.}} \textbf{10}, 84--86 (1963).

\bibitem{Glauber_1963_b}
R.~J. Glauber, \enquote{The quantum theory of optical coherence,} {\protect\JournalTitle{Phys. Rev.}} \textbf{130}, 2529--2539 (1963).

\bibitem{Kimble_1977_antBunch}
H.~J. Kimble, M.~Dagenais, and L.~Mandel, \enquote{Photon antibunching in resonance fluorescence,} {\protect\JournalTitle{Phys. Rev. Lett.}} \textbf{39}, 691--695 (1977).

\bibitem{Aspect_1981_Bell}
A.~Aspect, P.~Grangier, and G.~Roger, \enquote{Experimental tests of realistic local theories via bell's theorem,} {\protect\JournalTitle{Phys. Rev. Lett.}} \textbf{47}, 460--463 (1981).

\bibitem{dravins2012stellar}
D.~Dravins, S.~LeBohec, H.~Jensen, and P.~D. Nu{\~n}ez, \enquote{Stellar intensity interferometry: Prospects for sub-milliarcsecond optical imaging,} {\protect\JournalTitle{New Astronomy Reviews}} \textbf{56}, 143--167 (2012).

\bibitem{scarani2009security}
V.~Scarani, H.~Bechmann-Pasquinucci, N.~J. Cerf, \emph{et~al.}, \enquote{The security of practical quantum key distribution,} {\protect\JournalTitle{Reviews of modern physics}} \textbf{81}, 1301--1350 (2009).

\bibitem{lubin2022photon}
G.~Lubin, D.~Oron, U.~Rossman, \emph{et~al.}, \enquote{Photon correlations in spectroscopy and microscopy,} {\protect\JournalTitle{ACS photonics}} \textbf{9}, 2891--2904 (2022).

\bibitem{dutton_SPAD_arr}
N.~A. Dutton, I.~Gyongy, L.~Parmesan, \emph{et~al.}, \enquote{A {SPAD}-based {QVGA} image sensor for single-photon counting and quanta imaging,} {\protect\JournalTitle{IEEE Transactions on Electron Devices}} \textbf{63}, 189--196 (2015).

\bibitem{perenzoni_SPAD_arr}
M.~Perenzoni, N.~Massari, D.~Perenzoni, \emph{et~al.}, \enquote{A $160\times 120$ pixel analog-counting single-photon imager with time-gating and self-referenced column-parallel {A/D} conversion for fluorescence lifetime imaging,} {\protect\JournalTitle{IEEE Journal of Solid-State Circuits}} \textbf{51}, 155--167 (2015).

\bibitem{gyongy_SPAD_arr}
I.~Gyongy, N.~Calder, A.~Davies, \emph{et~al.}, \enquote{A $256\times256$, 100-kfps, 61\% fill-factor {SPAD} image sensor for time-resolved microscopy applications,} {\protect\JournalTitle{IEEE Transactions on Electron Devices}} \textbf{65}, 547--554 (2017).

\bibitem{ATLAS_1}
F.~M. Della~Rocca, E.~J. Sie, A.~T. Erdogan, \emph{et~al.}, \enquote{A 512$\times$ 512 {SPAD} laser speckle autocorrelation imager in stacked 65/40nm {CMOS},} in \emph{2024 IEEE Symposium on VLSI Technology and Circuits (VLSI Technology and Circuits),}  (IEEE, 2024), pp. 1--2.

\bibitem{ATLAS_2}
A.~Gorman, N.~Finlayson, A.~T. Erdogan, \emph{et~al.}, \enquote{{ATLAS}: a large array, on-chip compute {SPAD} camera for multispeckle diffuse correlation spectroscopy,} {\protect\JournalTitle{Biomed. Opt. Express}} \textbf{15}, 6499--6515 (2024).

\bibitem{SS1}
S.~Burri, Y.~Maruyama, X.~Michalet, \emph{et~al.}, \enquote{Architecture and applications of a high resolution gated {SPAD} image sensor,} {\protect\JournalTitle{Optics express}} \textbf{22}, 17573--17589 (2014).

\bibitem{SS2}
A.~C. Ulku, C.~Bruschini, I.~M. Antolovi{\'c}, \emph{et~al.}, \enquote{A 512$\times$ 512 {SPAD} image sensor with integrated gating for widefield {FLIM},} {\protect\JournalTitle{IEEE Journal of Selected Topics in Quantum Electronics}} \textbf{25}, 1--12 (2018).

\bibitem{Xavier_FLIM}
J.~T. Smith, A.~Rudkouskaya, S.~Gao, \emph{et~al.}, \enquote{In vitro and in vivo nir fluorescence lifetime imaging with a time-gated {SPAD} camera,} {\protect\JournalTitle{Optica}} \textbf{9}, 532--544 (2022).

\bibitem{Faccio_FLIM}
V.~Zickus, M.-L. Wu, K.~Morimoto, \emph{et~al.}, \enquote{Fluorescence lifetime imaging with a megapixel {SPAD} camera and neural network lifetime estimation,} {\protect\JournalTitle{Scientific Reports}} \textbf{10}, 20986 (2020).

\bibitem{defienne2021full}
H.~Defienne, J.~Zhao, E.~Charbon, and D.~Faccio, \enquote{Full-field quantum imaging with a single-photon avalanche diode camera,} {\protect\JournalTitle{Physical Review A}} \textbf{103}, 042608 (2021).

\bibitem{Gupta_3D}
F.~Gutierrez-Barragan, F.~Mu, A.~Ardelean, \emph{et~al.}, \enquote{Learned compressive representations for single-photon 3d imaging,} in \emph{Proceedings of the IEEE/CVF International Conference on Computer Vision (ICCV),}  (2023), pp. 10756--10766.

\bibitem{Dangelo_CPI}
G.~Massaro, P.~Mos, S.~Vasiukov, \emph{et~al.}, \enquote{Correlated-photon imaging at 10 volumetric images per second,} {\protect\JournalTitle{Scientific Reports}} \textbf{13}, 12813 (2023).

\bibitem{Dangelo_CPI_rev}
C.~Abbattista, L.~Amoruso, S.~Burri, \emph{et~al.}, \enquote{Towards quantum 3d imaging devices,} {\protect\JournalTitle{Applied Sciences}} \textbf{11}, 6414 (2021).

\bibitem{Faccio_SPAD_lidar}
J.~Zhao, A.~Lyons, A.~C. Ulku, \emph{et~al.}, \enquote{Light detection and ranging with entangled photons,} {\protect\JournalTitle{Optics Express}} \textbf{30}, 3675--3683 (2022).

\bibitem{Faccio_compImg}
Y.~Altmann, S.~McLaughlin, M.~J. Padgett, \emph{et~al.}, \enquote{Quantum-inspired computational imaging,} {\protect\JournalTitle{Science}} \textbf{361}, eaat2298 (2018).

\bibitem{Gupta_burst}
S.~Ma, S.~Gupta, A.~C. Ulku, \emph{et~al.}, \enquote{Quanta burst photography,} {\protect\JournalTitle{ACM Trans. Graph.}} \textbf{39} (2020).

\bibitem{Gupta_SodaCam}
V.~Sundar, A.~Ardelean, T.~Swedish, \emph{et~al.}, \enquote{Sodacam: Software-defined cameras via single-photon imaging,} in \emph{Proceedings of the IEEE/CVF International Conference on Computer Vision (ICCV),}  (2023), pp. 8165--8176.

\bibitem{Gur_quantCorr_Spad23}
G.~Lubin, R.~Tenne, I.~M. Antolovic, \emph{et~al.}, \enquote{Quantum correlation measurement with single photon avalanche diode arrays,} {\protect\JournalTitle{Opt. Express}} \textbf{27}, 32863--32882 (2019).

\bibitem{Uri_jointSR}
U.~Rossman, R.~Tenne, O.~Solomon, \emph{et~al.}, \enquote{Rapid quantum image scanning microscopy by joint sparse reconstruction,} {\protect\JournalTitle{Optica}} \textbf{6}, 1290--1296 (2019).

\bibitem{Lior_jointSR}
L.~M. Beck, A.~Shocher, U.~Rossman, \emph{et~al.}, \enquote{Improving correlation based super-resolution microscopy images through image fusion by self-supervised deep learning,} {\protect\JournalTitle{Optics Express}} \textbf{32}, 28195--28205 (2024).

\bibitem{STORM}
M.~J. Rust, M.~Bates, and X.~Zhuang, \enquote{Stochastic optical reconstruction microscopy ({STORM}) provides sub-diffraction-limit image resolution,} {\protect\JournalTitle{Nature methods}} \textbf{3}, 793 (2006).

\bibitem{PALM}
E.~Betzig, G.~H. Patterson, R.~Sougrat, \emph{et~al.}, \enquote{Imaging intracellular fluorescent proteins at nanometer resolution,} {\protect\JournalTitle{science}} \textbf{313}, 1642--1645 (2006).

\bibitem{SOFI}
T.~Dertinger, R.~Colyer, R.~Vogel, \emph{et~al.}, \enquote{Superresolution optical fluctuation imaging ({SOFI}),} {\protect\JournalTitle{Nano-Biotechnology for Biomedical and Diagnostic Research}} pp. 17--21 (2012).

\bibitem{Bruschini_microlenses}
C.~Bruschini, I.~M. Antolovic, F.~Zanella, \emph{et~al.}, \enquote{Challenges and prospects for multi-chip microlens imprints on front-side illuminated {SPAD} imagers,} {\protect\JournalTitle{Opt. Express}} \textbf{31}, 21935--21953 (2023).

\bibitem{Gur_heralded}
G.~Lubin, R.~Tenne, A.~C. Ulku, \emph{et~al.}, \enquote{Heralded spectroscopy reveals exciton--exciton correlations in single colloidal quantum dots,} {\protect\JournalTitle{Nano letters}} \textbf{21}, 6756--6763 (2021).

\end{thebibliography}

\end{document}